# Many-body interaction on near-field radiative heat transfer between two nanoparticles caused by proximate particle ensembles

Baokun Liu[a], Minggang Luo[a,b*], Junming Zhao[a,c*], Linhua Liu[d], Mauro Antezza[b,f*]

[a] School of Energy Science and Engineering, Harbin Institute of Technology, Harbin 150001, China
[b] Laboratoire Charles Coulomb (L2C) UMR 5221 CNRS-Universite de Montpellier, F-34095 Montpellier, France
[c] Key Laboratory of Aerospace Thermophysics, Ministry of Industry and Information Technology of China
[d] School of Energy and Power Engineering, Shandong University, Qingdao 266237, China
[f] Institut Universitaire de France, 1 rue Descartes, F-75231 Paris Cedex 05, France

**Abstract**

Near-field radiative heat transfer (NFRHT) has received growing attention because of its high intensity far beyond the Planck's black-body limit. Insertion of a third object in proximity of the two particles can significantly influence and manipulate its NFRHT. However, for the system composed of many particles, the effect of many-body interaction (MBI) on NFRHT between arbitrary two particles is still not well understood. In this work, the MBI is studied for two particles with three typical proximate ensembles: particle chain, plane and grating. With the increasing of proximate particle size, the MBI on NFRHT will experience a radical change from inhibition to enhancement. The polarizability of the proximate particle increases with particle radius, which enhances the interaction between the proximate particles and the main particle, and then results in enhancement of NFRHT between the main particles. When twisting the proximate particle ensemble, the proximate MBI accounts for a smooth and non-oscillated twisting angle dependence of NFRHT, different from the oscillation phenomenon of NFRHT for particle gratings. This work deepens the understanding of NFRHT in dense particulate systems.

* Corresponding author: luohit@126.com(Minggang Luo); jmzhao@hit.edu.cn(Junming Zhao); mauro.antezza@umontpellier.fr(Mauro Antezza)

# 1. Introduction

When the separation distance between two objects is comparable to or less than the characteristic thermal wavelength, the near-field radiative heat flux can exceed the Planckian blackbody limit heat flux due to the contribution of evanescent waves tunneling [1-5], which has been already experimentally verified [6-11]. Due to the super-Planckian effect, the near-field radiative heat transfer (NFRHT) has wide potential applications in thermpohotovoltaics, energy conversion devices and nanofabrication, to name a few [12-14]. With the insertion of other objects, the heat transfer between the existing two objects will inevitably be affected by the induced multiple scattering of the thermal excited evanescent waves, namely, the many-body interaction (MBI).

Many efforts have been spelt on the effect of MBI on the radiative heat transfer in particulate ensembles, recently. In the particulate ensemble, the particles often lie in the near field of each other. When considering heat transfer between two ensembles composed of many particles, it is the inter-ensemble MBI and intra-ensemble MBI that work together to affect the heat transfer [15], e.g., the heat flux can be significantly inhibited by the MBI when the separation between the two ensembles is less than characteristics thermal wavelength. The MBI effect on the radiative heat transfer between two clusters composed of many particles has been analyzed for both dielectric and metallic particles [16, 17]. For two dielectric particle clusters, the thermal conductance is inhibited but enhanced at frequencies close to the resonance frequency and the total thermal conductance between clusters is decreased because of the many-body interaction among particles [16]. While for metallic particle clusters, e.g., silver, the many-body interaction has a negligible effect on heat transfer at room temperature [17]. The inhibitive MBI effect on the radiative heat flux is also observed for the ensembles composed of many core-shell nanoparticles [18]. When considering the heat transfer between two particles extracted from the many-particle system, there are many particles lying nearby the two particles of interest, which will result in a complex multiple scattering. However, it is still unclear how the MBI induced by the nearby many particles affects the heat transfer between the two particles of interest.

The MBI effect on heat transfer in the three-body system has been analyzed already. The MBI induced by putting a third particle nearby the two particles in proximity will slightly inhibit the radiative heat flux [19]. A recent experimental research confirmed the inhibitive MBI effect on radiative heat flux for the planar structures [20]. However, the MBI can also significantly increase the radiative heat flux to around 10 times of magnitude by inserting a third particle at the center of the two particles [21-23]. This enhancement effect can be modulated by rotating the intermediate nanoparticle and can be explained by the coupling and mismatch of anisotropic local hyperbolic excitons. A recent study found that the addition of a third nanoparticle can lead to a great enhancement of the near-field radiative thermal conductance in an α-MoO3 three-body nanoparticles system due to the excitation of the localized hyperbolic polaritons [24]. Due to the negative differential thermal conductance effect the NFRHT between two Weyl semimetal nanoparticles can be greatly enhanced by exciting the localized plasmon and circular modes [25]. The radiative heat flux between two nanoparticles can be significantly amplified due to the presence of a proximate planar substrate supporting a surface resonance and the enhancement effect of dielectric particles is greater than that of metallic particles [26]. With the existence of a substrate placed in the near field, the NFRHT between two SiC nanoparticles can realize long-distance transport via propagating surface waves, which enhances the NFRHT by orders of magnitude at large distances [27]. The NFRHT between two α-MoO3 nanoparticles mediated by an α-MoO3 planar substrate due to the strong directionality of the localized hyperbolic polaritons in the nanoparticles and the volume-confined hyperbolic polaritons in the substrate [28]. In addition, when the nanoparticles are placed inside the cavity constructed by two parallel plates, the NFRHT will be modulated by the plates and a slower power-law decay is observed in the large distance regime [29]. By applying graphene to construct a nearby meta-surface whose optical properties are tunable, a giant modulation of the radiative heat flux between two particles can be achieved [30, 31]. It is reported that a near-field radiative thermal diode (NFRTD) based on two Weyl semimetal nanoparticles can be mediated by a Weyl semimetal planar substrate which is attributed to the strong coupling of the localized plasmon modes in the nanoparticles and nonreciprocal surface plasmon polaritons in the substrate [32]. Recently, Asheichyk and Krüger [33] found that a perfectly conducting nanowire nearby the two particles working as an excellent waveguide can transfer electromagnetic energy to far-field separation distance with almost no loss. It is interesting to know if the nearby particles can enhance the heat transfer between the two particles like a nearby substrate or perfectly conductive nanowire does.

For the two particles with a third object in proximity, it is worth to mention the effect of breaking symmetry of the system on the radiative heat transfer between the two particles. The symmetry-breaking is often applied to provide a platform for modulating the heat transfer for the micro-nanoscale thermal management devices, without the complex fabrication of nanostructure. When considering two anisotropic structures, the effect of symmetry breaking on heat transfer between the two structures by twisting a certain angle is analyzed. Due to the symmetry breaking by twisting one grating with an angle, a significant modulation of the near-field radiative heat flux between two gratings can be observed [34]. By tuning the external magnetic field and the twisting angle, the magnetoplasmonic modulation of heat transfer between two graphene gratings was studied [35]. A similar modulation of heat transfer between two magnetic Weyl semimetal slabs by twisting the intrinsic nonreciprocal surface modes was reported [36]. Most recently, low-symmetry Bravais crystal medium ($\beta$-$Ga_2O_3$) was reported as an excellent material for twist-induced modification due to its intrinsic shear effect [37]. The modulation factor of the rotated anisotropic α-$MoO_3$ nanoparticles could be up to ~12,000 due to the excitation of localized hyperbolic phonon polaritons (LHPPs) in different particle orientations [38]. When considering the grating composed of many particles, we previously studied the NFRHT between two twisted finite-size polar dielectric nanoparticle gratings with a focus on many-body interaction and we presented a characteristic oscillated dependence of the thermal conductance on the twisting angle [39]. It is interesting to know if symmetry breaking can be applied to modulate heat transfer between two particles by twisting the nearby structures.

When considering heat transfer between two particles, a modulation of the heat flux can be observed by twisting the hBN substrate with an angle, due to the directional energy of the hyperbolic surface phonon polaritons (HSPPs) in the hBN substrate [40]. A similar modulation of heat transfer between two particles is realized by twisting a nearby bilayer graphene grating [41]. It is interesting to know if we can still obtain a certain modulation of heat transfer between two particles but replace the nearby structures (e.g., hBN substrate or bilayer graphene grating) with an ensemble composed of common particles. It is still unclear if the oscillation of heat transfer on the twisting angle observed in Ref. [39] still exists for the heat transfer between two particles considering symmetry breaking by twisting.

To address the above missing points, the near-field radiative heat transfer between two particles with a many-particle ensemble lying in proximity is investigated by the coupled electric and magnetic dipole (CEMD) approach in the framework many-body radiative heat transfer theory, with a focus on the 'proximate' many-body interaction caused by the particles in proximity. This work is organized as follows. In Sec.II, the physical model of the considered two particles with a many-particle ensemble in proximity is also given. The formulas for the NFRHT and thermal conductance in the CEMD approach are presented in brief. In addition, the size effect of the proximate particle ensemble on the thermal conductance between the two particles is also presented. In Sec.III, the effect of the many-body interaction on heat transfer is analyzed, considering the two different influencing factors: (a) the filling ratio of the particulate ensemble and (b) the proximate particle size (radius). In addition, the effect of the spatial symmetry breaking of the proximate particle system on the heat transfer between two nanoparticles is also analyzed.

## 2. Theoretical background

### 2.2 Physical systems

We consider the heat transfer between two particles ('main particles') where there are a large number of particles ('proximate ensemble') in proximity, which is shown in Fig. 1. The proximate ensemble can be clarified into three kinds according to its structural characteristics: 1) one-dimensional (1D) particle chain, 2) two-dimensional (2D) particle plane and 3) 2D particle grating. Note that in Fig. 1, the 'particle plane' refers to the case 2) that the lattice spacings in the two lateral directions are identical to each other, and the 'particle grating' refers to the case 3) that the two abovementioned lattice spaces are quite different from each other.

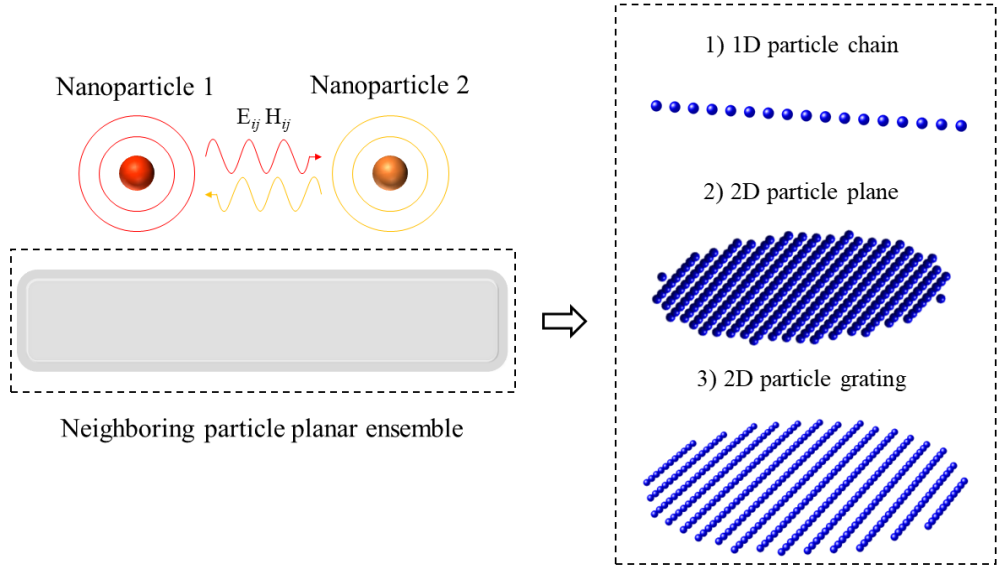

Fig. 1  Schematic diagram of the nanoparticle system.

All the necessary configuration parameters are shown in Fig. 2. The two main nanoparticles are in red with a separation $l$ and radius $R'$. The parallel and vertical lattice spacings of the proximate ensemble (particles radius $R$) are denoted by $h_1$ and $h_2$, respectively. We fixed the central particle of the proximate ensemble (the one in the blue shadow shown in Fig. 2) below the midpoint of the connecting line of the two main particles. The separation distance between the main particles and the proximate ensemble is $d$. When considering twisting the lower proximate ensemble, the relative twisting angle between the two particles and the neighboring ensemble is $\theta$.

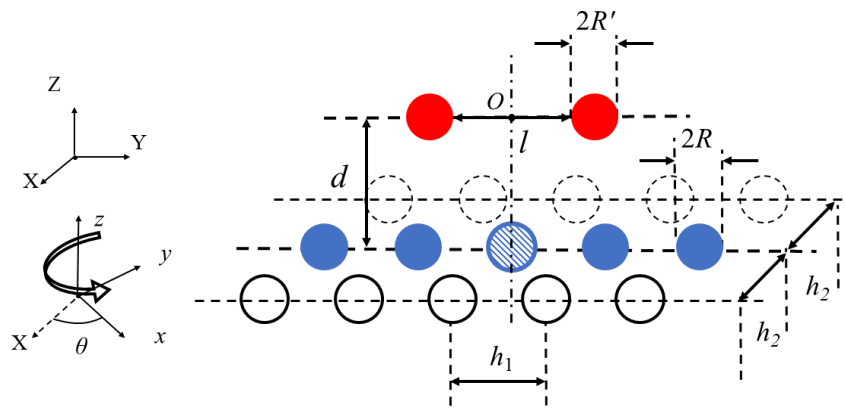

Fig. 2  Simplified configuration of the neighboring particle system.

## 2.2 Formulas

In the framework of the many-body radiative heat transfer theory for near-field radiative heat flux between particles, the radiative thermal conductance ($G$) between two main nanoparticles is defined as [17]

$$G = \lim_{\delta T \to 0} \frac{P_{1 \leftrightarrow 2}}{\delta T}, \tag{1}$$

where $\delta T$ is the temperature difference between the emitting particle and the absorbing particle. The exchanged radiative power between the main nanoparticle 1 and nanoparticle 2 is [19, 21]

$$\begin{aligned} P_{1 \leftrightarrow 2} &= P_{1 \to 2} - P_{2 \to 1} \\ &= 3 \int_0^{+\infty} \frac{d\omega}{2\pi} (\Theta(\omega, T_1) - \Theta(\omega, T_2)) \mathcal{T}_{1,2}(\omega) \end{aligned}, \tag{2}$$

where the $P_{j \to i}$ denotes power absorbed by particle $i$ and radiated by particle $j$, which can be calculated by

$$P_{j \to i} = 3 \int_0^{+\infty} \frac{d\omega}{2\pi} \Theta(\omega, T_j) \mathcal{T}_{i,j}(\omega), \tag{3}$$

where $\Theta(\omega, T_j)$ is the mean energy of a harmonic Planck oscillator and $\mathcal{T}_{i,j}(\omega)$ is the transmission coefficient between the $j$th and the $i$th particle, which is given as follows.

$$\begin{aligned} \mathcal{T}_{i,j}(\omega) = \frac{4}{3} k^4 [&\text{Im}(\chi_E^i) \text{Im}(\chi_E^j) \text{Tr}(G_{ij}^{EE} G_{ij}^{EE\dagger}) \\ &+ \text{Im}(\chi_E^i) \text{Im}(\chi_H^j) \text{Tr}(G_{ij}^{EM} G_{ij}^{EM\dagger}) \\ &+ \text{Im}(\chi_H^i) \text{Im}(\chi_E^j) \text{Tr}(G_{ij}^{ME} G_{ij}^{ME\dagger}) \\ &+ \text{Im}(\chi_H^i) \text{Im}(\chi_H^j) \text{Tr}(G_{ij}^{MM} G_{ij}^{MM\dagger})] \end{aligned}, \tag{4}$$

where parameters $\chi_E = \alpha_E - \frac{ik^3}{6\pi} |\alpha_E|^2$ and $\chi_H = \alpha_H - \frac{ik^3}{6\pi} |\alpha_H|^2$. The $G_{ij}^{\nu\tau}$ ($\nu, \tau = E$ or $M$) is the Green's function and it can be calculated in the compact form [42]

$$\mathbb{G}_{ij} = \begin{pmatrix} \mu_0 \omega^2 G_{ij}^{EE} & \mu_0 \omega G_{ij}^{EM} \\ k\omega^2 G_{ij}^{ME} & k^2 G_{ij}^{MM} \end{pmatrix}. \tag{5}$$

The $\mathbb{G}_{ij}$ is the solution of the following equation

$$\begin{pmatrix} 0 & \mathbb{G}_{12} & \cdots & \mathbb{G}_{1N} \\ \mathbb{G}_{21} & 0 & \ddots & \vdots \\ \vdots & \vdots & \ddots & \mathbb{G}_{(N-1)N} \\ \mathbb{G}_{N1} & \mathbb{G}_{N2} & \cdots & 0 \end{pmatrix} = \begin{pmatrix} 0 & \mathbb{G}_{0,12} & \cdots & \mathbb{G}_{0,1N} \\ \mathbb{G}_{0,21} & 0 & \ddots & \vdots \\ \vdots & \vdots & \ddots & \mathbb{G}_{0,(N-1)N} \\ \mathbb{G}_{0,N1} & \mathbb{G}_{0,N2} & \cdots & 0 \end{pmatrix} \mathbf{A}^{-1}, \tag{6}$$

where the Green's function in free space is

$$\mathbb{G}_{0,ij} = \begin{pmatrix} \mu_0 \omega^2 G_{0,ij}^{EE} & \mu_0 \omega G_{0,ij}^{EM} \\ k\omega^2 G_{0,ij}^{ME} & k^2 G_{0,ij}^{MM} \end{pmatrix} \quad (7)$$

where the free space Green's function connecting two nanoparticles at $\mathbf{r}_i$ and $\mathbf{r}_j$ are defined as follows,

$$G_{0,ij}^{EE} = \frac{e^{ikr}}{4\pi r}\left[(1+\frac{ikr-1}{k^2-r^2})\mathbf{I}_3 + \frac{3-3ikr-k^2r^2}{k^2r^2}\hat{\mathbf{r}} \otimes \hat{\mathbf{r}}\right], \quad (8)$$

$$G_{0,ij}^{ME} = \frac{e^{ikr}}{4\pi r}\left(1-\frac{1}{ikr}\right)\begin{pmatrix} 0 & -\hat{r}_z & \hat{r}_y \\ \hat{r}_z & 0 & -\hat{r}_x \\ -\hat{r}_y & \hat{r}_x & 0 \end{pmatrix}, \quad (9)$$

where $\mathbf{I}_3$ is a 3×3 identity matrix and $r$ is the magnitude of the separation vector $\mathbf{r}=\mathbf{r}_i-\mathbf{r}_j$. $\hat{\mathbf{r}}$ is the unit vector $\mathbf{r}/r$. $\otimes$ denotes the outer product of vectors. The matrix $\mathbf{A}$ including many-body interaction is defined as,

$$\mathbf{A} = \mathbf{I}_{6N} - \begin{pmatrix} 0 & \boldsymbol{\alpha}_1 \mathbb{G}_{0,12} & \cdots & \boldsymbol{\alpha}_1 \mathbb{G}_{0,1N} \\ \boldsymbol{\alpha}_2 \mathbb{G}_{0,21} & 0 & \ddots & \vdots \\ \vdots & \vdots & \ddots & \boldsymbol{\alpha}_{N-1} \mathbb{G}_{0,(N-1)N} \\ \boldsymbol{\alpha}_N \mathbb{G}_{0,N1} & \cdots & \boldsymbol{\alpha}_N \mathbb{G}_{0,N(N-1)} & 0 \end{pmatrix}, \quad (10)$$

where $\boldsymbol{\alpha}_i$ is defined as

$$\boldsymbol{\alpha}_i = \begin{pmatrix} \varepsilon_0 \alpha_E^i \mathbf{I}_3 & 0 \\ 0 & \alpha_H^i \mathbf{I}_3 \end{pmatrix}. \quad (11)$$

For the isotropic spherical particle, electric and magnetic dipole polarizabilities are $\alpha_E = \frac{i6\pi}{k^3}a_1$ and $\alpha_H = \frac{i6\pi}{k^3}b_1$, respectively, where $a_1$ and $b_1$ are the first-order Lorenz-Mie scattering coefficients, respectively, they are defined as [43]

$$a_1 = \frac{\varepsilon j_1(y)[xj_1(x)]' - j_1(x)[yj_1(y)]'}{\varepsilon j_1(y)[xh_1^{(1)}(x)]' - h_1^{(1)}(x)[yj_1(y)]'}, \quad (12)$$

$$b_1 = \frac{j_1(y)[xj_1(x)]' - j_1(x)[yj_1(y)]'}{j_1(y)[xh_1^{(1)}(x)]' - h_1^{(1)}(x)[yj_1(y)]'}, \quad (13)$$

where $x = kR'$ and $y = \sqrt{\varepsilon}kR'$, $k$ is the wave vector, $j_1(x) = \sin(x)/x^2 - \cos(x)/x$ and $h_1^{(1)}(x) = e^{ix}(1/ix^2 - 1/x)$ are Bessel functions and spherical Hankel functions, respectively. $\varepsilon$ is the dielectric permittivity. All particles are SiC particles in this work and the dielectric function of SiC can be described by the Drude-Lorentz model [44],

$$\varepsilon(\omega) = \varepsilon_\infty \frac{\omega^2 - \omega_l^2 + i\Gamma\omega}{\omega^2 - \omega_t^2 + i\Gamma\omega}, \qquad (14)$$

where $\omega_t = 1.495 \times 10^{14}\,\mathrm{rad \cdot s^{-1}}$, $\omega_l = 1.827 \times 10^{14}\,\mathrm{rad \cdot s^{-1}}$, $\varepsilon_\infty = 6.7$ and $\Gamma = 0.9 \times 10^{12}\,\mathrm{rad \cdot s^{-1}}$.

## 2.3 Boundary effect

As reported in our previous work [39], the size of the particle ensemble can usually affect the heat transfer, namely the size effect or the boundary effect. To avoid the size effect on the heat transfer and to approach a convergent result of heat transfer, we show the dependence of the thermal conductance between the main particles on the lateral length of the proximate ensemble (or the number of particles in the parallel direction $N$) in Fig. 3 for two kinds of proximate ensembles: (a) 1D particle chain, and (b) 2D particle plane. The conductance is calculated at 300 K. When increasing the particle number $N$, the thermal conductance ($G$) between two nanoparticles varies sharply at the very beginning. Then, the effect of the particles located in the boundary on the $G$ becomes less and less important. To obtain a stable and convergent result concerning the thermal conductance $G$ between the main particles with considering proximate particles, the number of particles in the proximate ensemble should be large enough, where the boundary has no effect on the heat transfer and the proximate structure can be regarded as infinite.

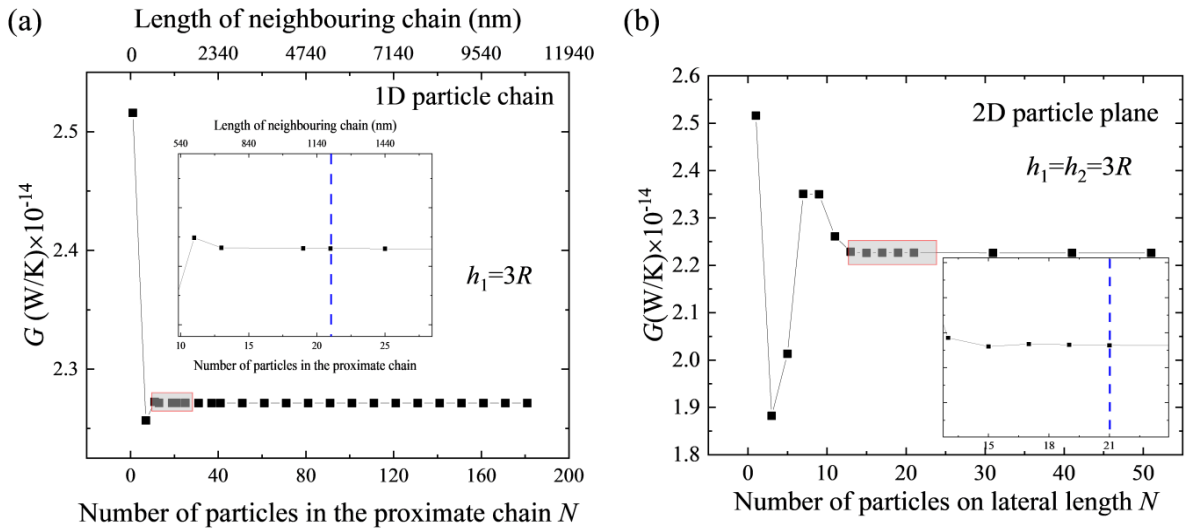

Fig. 3    Thermal conductance (*G*) as a function of the particle number of neighboring particle systems. (a) 1D particle chain, (b) 2D particle plane. *R*=20nm, *R'*=20nm, *l*=100nm and *d*=100nm.

## 3. Results and discussions

In this section, the many-body interaction (MBI) effect on the near-field radiative heat transfer (NFRHT) caused by the particle ensemble standing in proximity is investigated at room temperature (300K), namely the 'proximate' MBI. Three influencing factors are considered, i.e., the particle filling ratio of the proximate particle ensemble, the size of the particle and the symmetry-break caused by the twisting of the proximate particle ensemble. The whole nanoparticle system is in vacuum. Note that the minimum separation between particles is no less than three times the particle radius, which makes the dipole approximation valid [45, 46]. To avoid the boundary effect, we take $N = 21$ particles for each edge in the 1D and 2D proximate particle ensembles.

### 3.1 Effect of particle filling ratio of neighboring ensemble on NFRHT

As can be seen in Fig. 2, the particle filling ratio of the proximate structure can be modified by changing the particle spacing (i.e., $h_1$ and $h_2$) in different directions. The dependence of thermal conductance *G* between the two main nanoparticles on the dimensionless vertical particle spacing $h_2$ of 2D neighboring particle ensemble is shown in Fig. 4. The conductance between two main particles without any proximate particles is also added for reference, shown as the red dash line. The dot lines are for the thermal conductance of the cases considering only the existence of the 1D chain (shown as the blue particle chain in Fig. 2) in most proximity. We can see that the proximate system inhibits the NFRHT due to the many-body interaction, which is consistent with the reported inhibitive many-body interaction on thermal conductance distribution along a particle chain caused by the proximate particle chain in our previous studies [16, 47]. The inhibition decreases significantly with increasing separation distance *d*. As shown in Fig. 4, when increasing $h_2$, all the curves converge to the thermal conductance ($G_s$) corresponding to the limit situation that only the particle chain standing solely right beneath the two main particles exists in proximity. When $h_2$ is larger enough (ten times the particle radius), the proximate MBI caused by the proximate 2D particle ensemble on the NFRHT can be simplified as that caused by only a proximate 1D particle chain, which will hence significantly reduce the computational efforts from the consumption point of view. It is also worthwhile mentioning that the $G_s$ increases when increasing the separation between the main particles and neighboring particle ensemble, which is due to the less and less important inhibitive MBI effect.

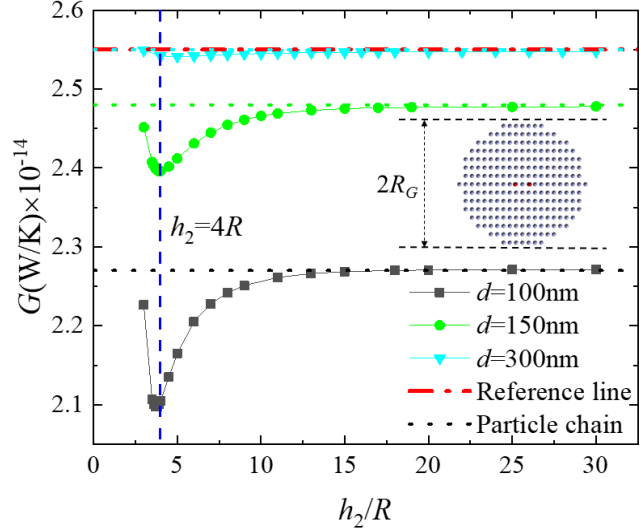

Fig. 4　Thermal conductance ($G$) as a function of normalized parallel spacing $h_2/R$. Three different separation distances are considered, $d$=100nm, $d$=150nm, and $d$=300nm, respectively. $R$=20nm, $R'$=20nm, $l$=100nm, $h_1$=3$R$ and $R_G$=600nm. Red dash line refers to the result without considering any neighboring proximate particles. The dot lines are the results of the case that the neighboring particle chain in most proximity stands solely.

Before numerical simulations, we may predict qualitatively and generally that the denser the proximate structure is, the stronger the proximate MBI effect will be. As shown in Fig. 4, it is obvious that the thermal conductance $G$ between the two main particles does not vary monotonically with increasing $h_2$. When increasing $h_2$, there is an extremum value of the thermal conductance $G$, which occurs roughly at $h_2$=4$R$. The diagrams of the considered three cases from the top viewpoint are given Fig. 5. It seems the MBI reaches its peak in a relatively dilute case ($h_2$=4$R$) rather than a much denser case (e.g., $h_2$=3$R$), which is a little bit beyond imagination and will be explained from the spectrum of the thermal conductance point of view. The spectrum at around the resonance of the thermal conductance $G_\omega$ between the two main nanoparticles with proximate particle ensembles having different $h_2$ (i.e., $h_2$=3$R$, 4$R$ and 20$R$) at separation $d$=150nm is given in Fig. 6. The peaks are due to the localized surface phonon polariton resonance frequency of spherical SiC particles, which is determined by Fröhlich condition Re[$\varepsilon(\omega)$]+2=0. As shown in Fig. 6, changing $h_2$ has no effect on the position of the peak of the thermal conductance spectrum, but will only affect the absolute value of the peak. It is because the many-body interaction between the proximate particle ensemble and the main particles will not change the spectrum trend of the thermal conductance but slightly affect the value itself [15].

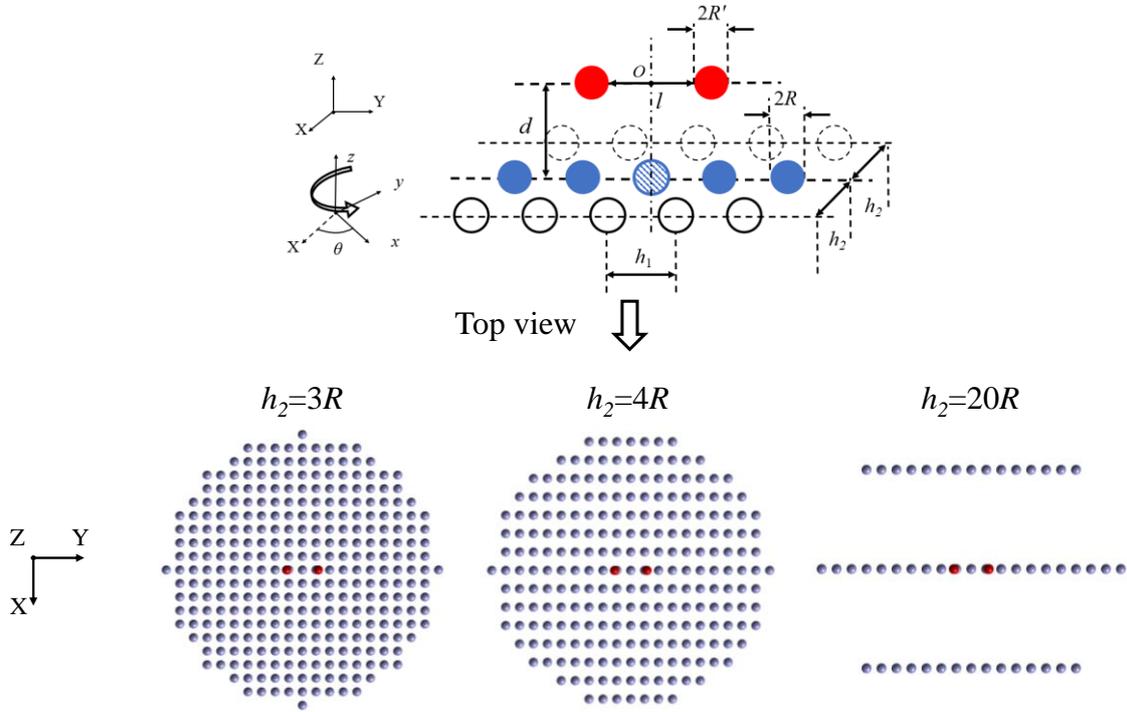

Fig. 5 Top view of the several representative actual particle systems. Red spheres are the two main nanoparticles and others are neighboring particles.

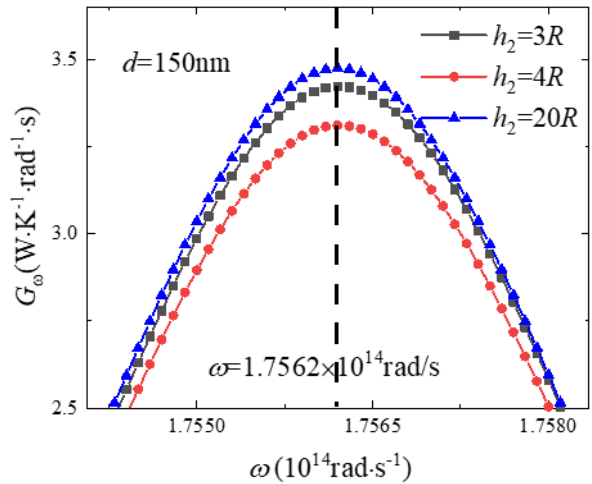

Fig. 6 Spectral thermal conductance ($G_\omega$) between two main SiC nanoparticles with several $h_2$ at $d = 150$nm.

The effect of changing the particle spacing in the parallel direction $h_1$ on the NFRHT between the two main particles is also analyzed. The dependence of the thermal conductance ($G$) on the dimensionless parallel spacing $h_1/R$ is shown in Fig. 7. As expected, $G$ in all cases will converge with increasing $h_1$, regardless of how large the separation $d$ is. When $h_2$ is very large (e.g., $h_2=20R$), the thermal conductance considering the proximate particle ensemble is nearly identical to that of the cases with the particle chain in the most proximity solely (denoted by red dotted lines). The smaller the $h_2$ is, the greater the deviation between the thermal conductance results of the cases with the proximate particle ensemble and that of the cases with a particle chain in most proximity solely will be. It shows that when $h_2$ is large enough, the neighboring 2D particle ensemble can be simplified as the nanoparticle chain in the most proximity.

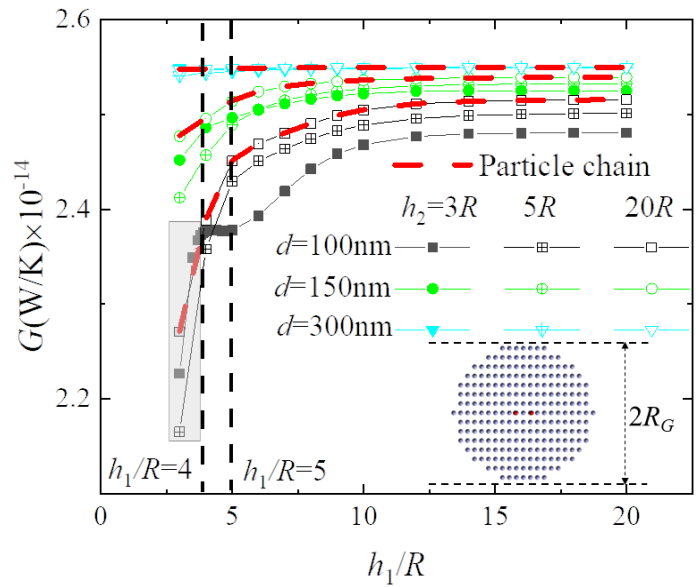

Fig. 7  Thermal conductance ($G$) between two main SiC nanoparticles as a function of normalized parallel spacing $h_1/R$. $R=20$nm, $R'=20$nm, $l=100$nm and $R_G=600$nm. Red dotted lines are the results of the cases of particle chains.

The $G$ between the two main particles increases monotonously with $h_1$, which is different from the trend observed in Fig. 4. It shows that the effect of proximate many-body interaction on the NFRHT is related to the relative orientation between the proximate particle ensemble and the main particles. As can be seen from the gray region in Fig. 7, when $h_1$ is very small, the inhibitive effect on the thermal conductance decreases when $h_2$ varies from $5R$ to $3R$ and further $20R$. It is consistent with the observations in Fig. 4, which indicates that the case of $h_2 = 5R$ has stronger inhibitive MBI on NFRHT than the case of $h_2 = 3R$ does. When $h_1$ becomes large enough (around $4.5R$), this special phenomenon will disappear and the result will become commonsensible, namely, a denser proximate structure has a stronger inhibitive MBI effect. In addition, it can be observed that there is an odd region where $G$ remains constant between $h_1/R = 4$ and 5 with $d = 100$nm. The spectral thermal conductance $G_\omega$ of several representative $h_1$ is shown in Fig. 8. Obviously, the results of $h_1/R=4.2$ and 4.8 are almost the same, which leads to the invariance of $G$ as shown in Fig. 7.

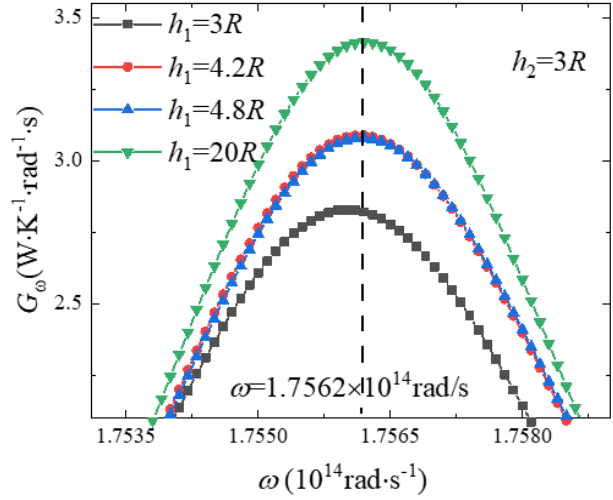

Fig. 8　Spectral thermal conductance ($G_\omega$) between two main SiC nanoparticles with different $h_1$ at $d$=100nm.

### 3.2 Effect of particle radius of neighboring ensemble on NFRHT

According to **Section 3.1**, the neighboring particle system can be represented by one particle chain when one direction particle spacing $h_1$ or $h_2$ is very large. Hence, the results of the case considering a proximate chain can be regarded as that of the case considering a proximate particle system with $h_2 = 600$ nm. The dependence of the thermal conductance ($G$) between the two main particles with a neighboring chain on the proximate particle radius ($R$) is shown in Fig. 9. As shown, the thermal conductance changes with increasing the proximate particle size $R$ non-monotonically. Specifically, when increasing $R$, the thermal conductance decreases at first and then increases again, which correspondingly means that the inhibitive MBI caused by the proximate chain increases at first and then decreases. When increasing the $R$ further, the MBI effect on NFRHT will experience a radical change from inhibition to enhancement on NFRHT. The previous work reported that by inserting a third particle in the center position of the connection line of the two particles can significantly enhance the NFRHT by about 1 order of magnitude [21], however putting a third particle nearby the two particles in proximity rather than in the center of its connection line can inhibit the NFRHT slightly [19, 47]. It is worth mentioning that the unique enhancement on NFRHT caused by the proximate particle chain has never been reported before. When increasing the radius sufficiently, the proximate particles are so big as compared to the two main particles of interest that the effect of particles in most proximity is similar to that of a nearby slab which the radiative heat flux between two nanoparticles can be significantly enhanced by inserting a proximate planar substrate supporting a surface resonance [26, 27].

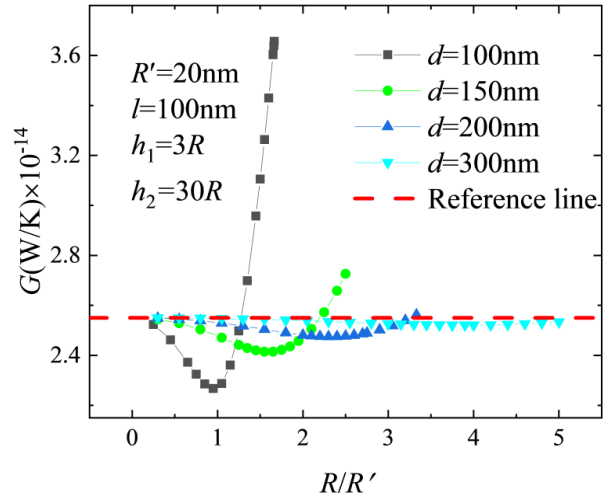

Fig. 9 Thermal conductance ($G$) between two main SiC nanoparticles as a function of $R$ with a neighboring chain. The results without the existence of neighboring nanoparticles are added for reference.

To help for understanding, the spectral thermal conductance and spectral polarizability of different sizes of nanoparticles are shown in Fig. 10 (a) and (b), respectively. As can be seen, the spectral thermal conductance $G_\omega$ decreases at first and then increases with the increase of $R$, which lead to an identical trend of thermal conductance $G$. As shown in Fig. 10 (b), the particle size $R$ affects the peak of spectral polarizability of particle and the larger the $R$, the larger the peak value. The increased proximate particle polarizability caused by increasing particle radius will affect the many-body interaction in the particle ensembles characterized by the interaction matrix **A** (see the polarizability of all particles in Eq. (10)), which hence results in an enhancement effect on NFRHT.

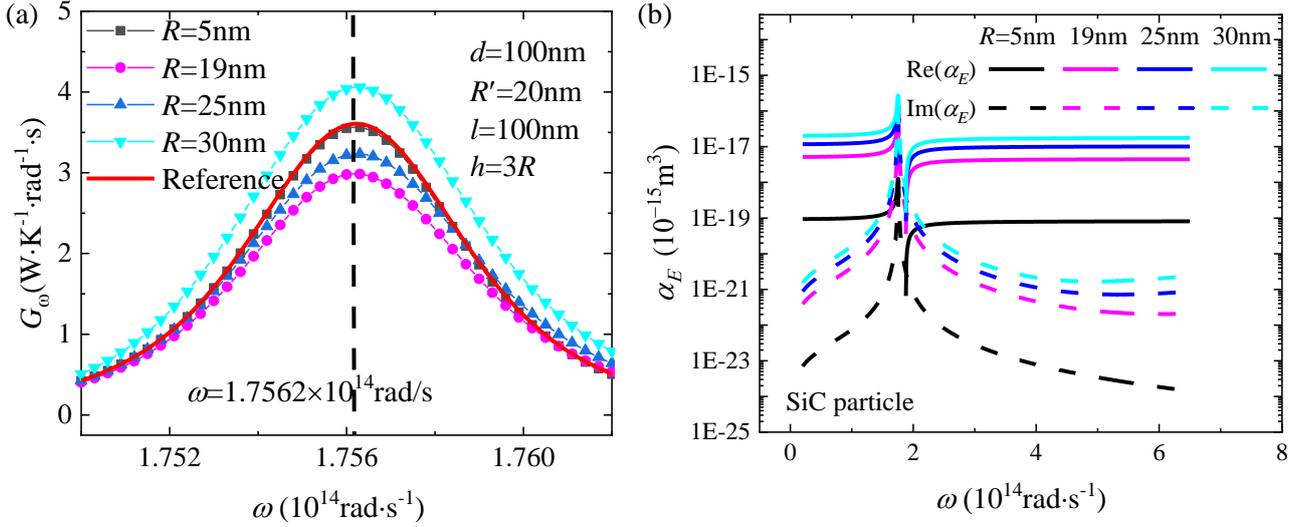

Fig. 10  Spectral parameters for the NFRHT between particles: (a) spectral thermal conductance between two main SiC nanoparticles with different *R*, (b) spectral polarizability of the nanoparticles with different sizes.

The abovementioned discussion concerning the Fig. 9 is for a very large $h_2$, so that the proximate particle ensemble can be seen as a proximate chain. Here, we decrease $h_2$ to study the effect of particle size in a 2D proximate geometry, $h_2 = 10R$ and $h_2 = 3R$, respectively. The dependence of the thermal conductance (*G*) between the two main particles on the neighboring particle radius (*R*) is shown in Fig. 11. The variation trends of the results with different $h_2$ are consistent with each other and the reason can be found in the results of spectral thermal conductance $G_\omega$ given in Fig. 12. It can be observed that as *R* increases, the peak value of $G_\omega$ decreases at first and then increases, which can be attributed to the many-body interaction, The variation of particle size has a significant influence on the peak value, but a very limited impact on the peak position. Obviously, the results of $h_2 = 30R$ are essentially identical to that of $h_2 = 10R$, and the results for $h_2 = 3R$ are a little different from the other two. It also proves that the 2D particle system can be approximated by a 1D particle chain when $h_2$ is sufficiently large, which has already been concluded in **Sec. 3.1**. As can be seen that the inhibition effect is generally strongest when $h_2 = 3R$, which indicates that the influence of the filling ratio of proximate systems is dominant, namely, the denser the system is, the stronger the inhibition effect is. It is also worthwhile mentioning that when *R* is very small, we can see no proximate MBI effect on NFRHT for all the considered cases at all because the proximate particle is too small to be seen.

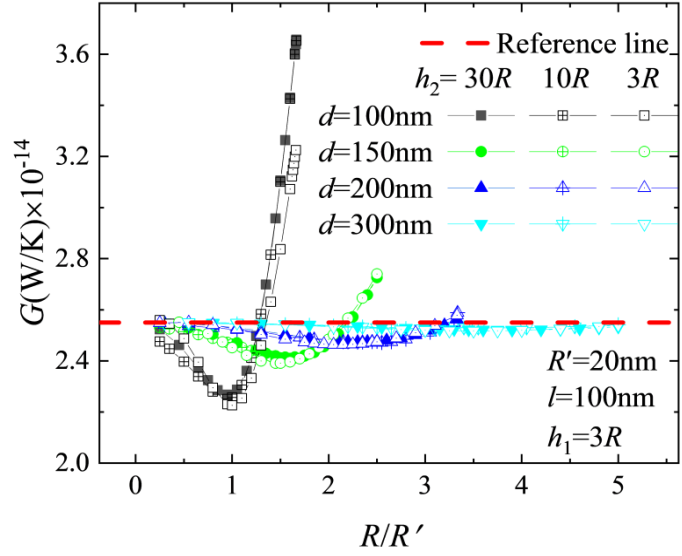

Fig. 11    Thermal conductance ($G$) between two main SiC nanoparticles as a function of $R$ with a neighboring ensemble. Reference line refers to the results without the existence of neighboring nanoparticles.

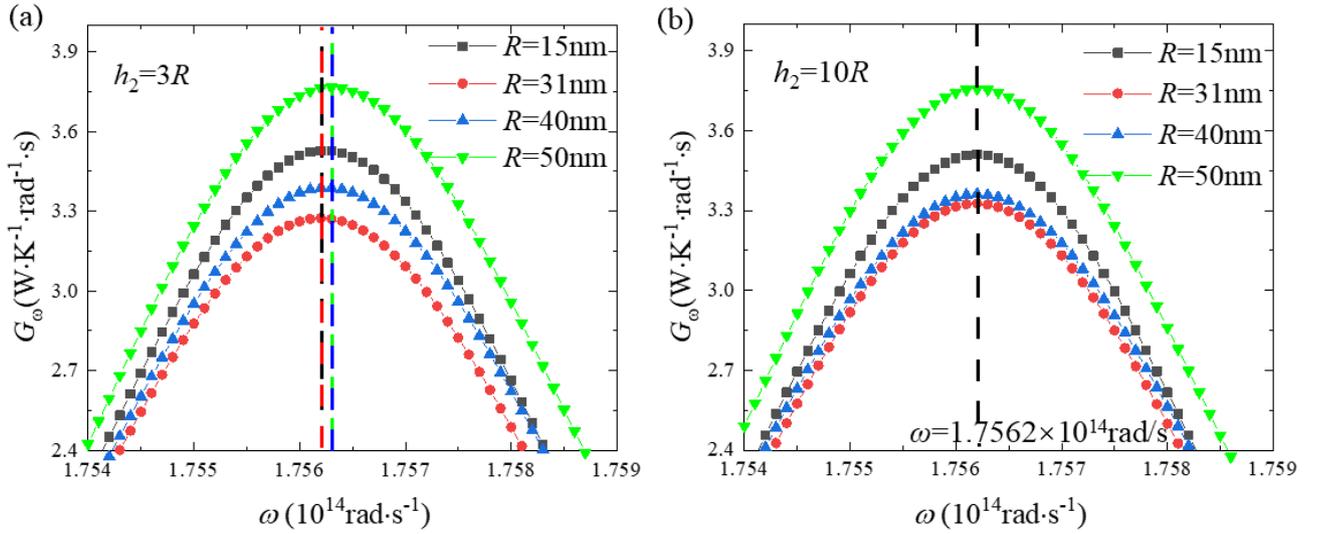

Fig. 12    Spectral thermal conductance ($G_\omega$) between two main SiC nanoparticles with different $R$ for particle planes and particle gratings at $d = 100$ nm.

**3.3 Effect of spatial rotation symmetry-break of neighboring ensemble on NFRHT**

The twisting- and distance-dependent coupling between the two nanoparticle gratings have an important effect on the NFRHT due to the breaking of spatial symmetry [39]. It is still unclear how symmetry breaking affects the NFRHT between two main particles with a proximate ensemble, which is the investigation focus of this section. The effect of the twisting angle of the neighboring particle ensemble on the radiative heat transfer between the two main SiC particles is analyzed. The dependence of the thermal conductance between the two nanoparticles and the modulation ratio $\varphi = G(\theta)/G(0°)$ on the twisting angle $\theta$ is shown in Fig. 13, respectively. The thermal conductance between two particles without any proximate particles is also added in Fig. 13 (a) for reference, symbolled by the red dash reference. All the considered proximate particle structures are rotationally symmetrical about the origin of the coordinate and the twisting angle $\theta$ ranges from 0 to 90 degrees. When the proximate particle ensemble exists, the thermal conductance is no greater than that of the two particles without any proximate particles. For a large separation (e.g., 300 nm), the proximate MBI is negligible. While for a short separation (e.g., 100 nm and 150 nm), the proximate MBI caused by the neighboring particles inhibits the NFRHT under different twisting angle $\theta$. As expected, the results of particle chain ($h_2 = 600$ nm) and particle grating ($h_2 = 200$ nm, $h_2 > h_1$) are almost identical to each other, which also confirms the conclusion in **Sec. 3.1**.

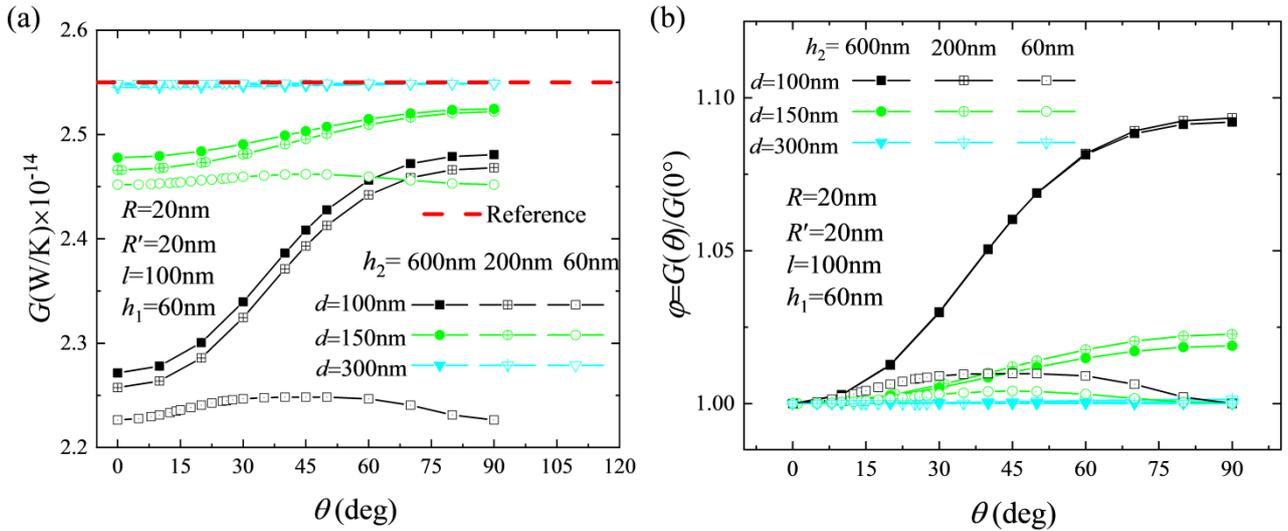

Fig. 13  Thermal conductance (a) and modulation ratio (b) between two SiC nanoparticles as a function of twisting angle $\theta$. Reference line refers to the results without the existence of a neighboring system.

For the cases that $h_1 < h_2$, the thermal conductance $G$ increases monotonically with the twisting angle $\theta$, although the increasing rate of the thermal conductance $G$ is not a constant. For the case that $h_2 = h_1 = 60$ nm, the dependence of the thermal conductance on the twisting angle is symmetric about a special twisting angle (around $\theta = 45$ degrees), as shown in Fig. 13 (a). To help the understanding of the twisting-proximate-particles NFRHT, we show the geometry diagram of the proximate particle ensemble in the twisting process, as shown in Fig. 14. After twisting the proximate particle by a certain angle, the neighboring particles will move away from their original positions where the proximate particles lie directly below the main particles. In the considered twisted angle range (0 to 90 degrees for $h_1<h_2$ and 0 to 45 degrees for $h_1=h_2$), the twisting of proximate structured particles will result in an increasing distance between the proximate particles and main particles and thus a less and less important proximate MBI effect on NFRHT. It is worthwhile mentioning that when $h_1= h_2$, the proximate particle ensemble is rotationally symmetric about the coordinate origin by about 90 degrees, rather the 180 degrees for the other two particle ensembles (i.e., particle chain and particle grating).

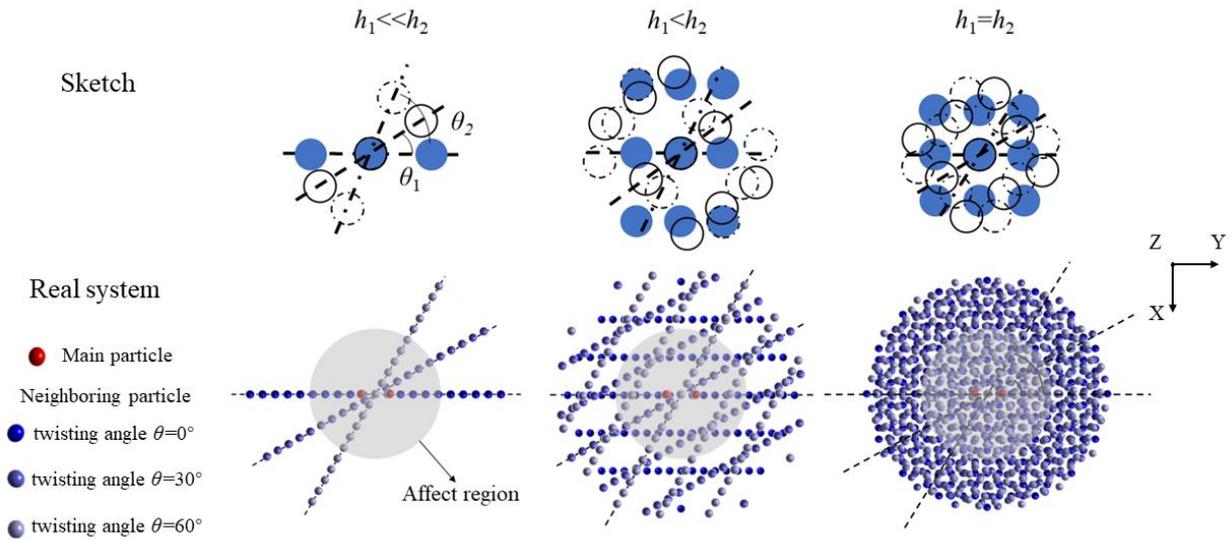

Fig. 14   Top view of twisting neighboring particle ensembles with different angle $\theta$.

In Fig. 13 (b), the dependence of the modulation ratio $\varphi = G(\theta)/G(0°)$ on the twisting angle $\theta$ is shown. For the cases that $h_1<h_2$ (i.e., $h_1= 60$ nm, $h_2=200$ nm and 600 nm), the modulation ratio range increases with decreasing the separation between proximate particles and main particles. When the separation is very large, the NFRHT is insensitive to the relative orientation anymore, the twisting of the proximate particles will not affect the thermal conductance, which is consistent with the observation for NFRHT between two particle clusters [17]. When the separation is large enough, the effect of the orientation of the particle cluster on NFRHT becomes less important. For the cases that $h_1= h_2 = 60$ nm, the modulation ratio $\varphi = G(\theta)/G(0°)$ is symmetric about the twisting angle of 45 degrees, which results from its geometrically rotational symmetry by about 90 degrees. The modulation ratio range is narrow as compared to that of the corresponding cases that $h_1 \neq h_2$. Because when $h_1 = h_2 = 60$ nm, the particles in the proximate ensemble are close to each other, no matter what the twisting angle is, the proximate MBI is still strong. When increasing the twisted angle, the proximate MBI will not change too much because the separation between the main particles and proximate particles does not change too much. However, for the cases that $h_1 \neq h_2$, the proximate MBI significantly changes when twisting the proximate particle by a certain angle, because the separation between the main particles and proximate particles changes a lot.

The spectral thermal conductance $G_\omega$ with different twisting angle $\theta$ are shown in Fig. 15. The separation between the proximate particle ensemble and main particles $d$ is 100 nm. $h_2 = 600$ nm ($30R$) and 200 nm ($10R$), respectively. The dependence of the spectral thermal conductance for different twisting angles is identical to each other for the two lattice spacings, i.e., $h_2 = 30R$ and $10R$. As can be seen that twisting of the proximate particle ensemble only affects the value of the peak of the $G_\omega$ and has no effect on the peak position. The larger the twisting angle is, the weaker the inhibition caused by the proximate particle ensemble on NFRHT is. With the lattice spacing $h_2$ increasing, the twisting-dependence of thermal conductance tends to converge. In addition, the dependence of $G$ on the twisting angle in this work is smooth and monotonic, and no oscillation exists, which is quite different from the observation in Ref. [39]. The proximate MBI will not bring an oscillation of NFRHT when twisting the proximate particle ensemble. However, the intra-ensemble and inter-ensemble MBI account for such oscillation phenomenon of NFRHT reported for particle gratings [15]. It can be noted that a second peak of $\varphi$ can be observed in Fig. 15(b) at narrow angles, which is caused by the many-body interaction.

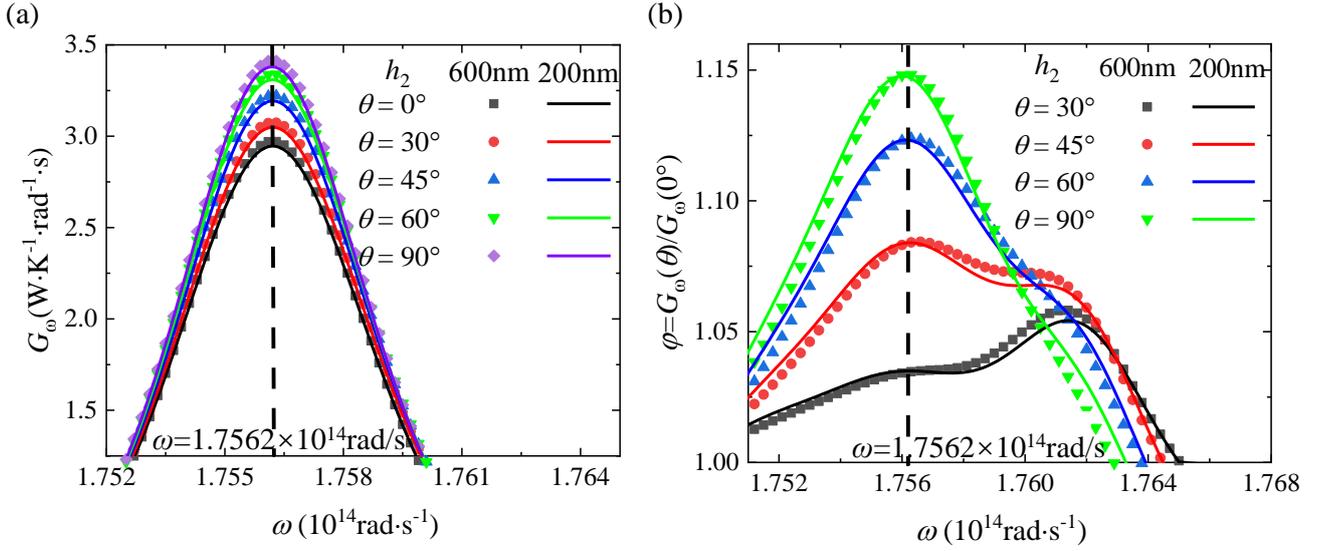

Fig. 15 Spectral parameters for heat transfer between two main SiC nanoparticles with different twisting angle $\theta$. (a) thermal conductance ($G_\omega$); (b) $\varphi = G_\omega(\theta)/G_\omega(0°)$. $d$=100 nm, $l$=100nm, $R$=20nm, $R'$=20nm and $h_1$=60nm.

## 4. Conclusion

The effect of the proximate (neighboring) nanoparticle ensembles on the NFRHT between two nanoparticles is investigated by using the many-body radiative heat transfer theory. In general, the proximate particle system in proximity will inhibit the NFRHT between the two particles due to the many-body interaction, which is consistent with the reported inhibitive many-body interaction on thermal conductance distribution along a particle chain caused by the proximate particle chain in our previous studies. The thermal conductance changes with increasing the proximate particle size $R$ non-monotonically. When increasing $R$, the MBI effect on NFRHT will experience a radical change from inhibition to enhancement on NFRHT. It is worth mentioning that the unique enhancement on NFRHT caused by the proximate particle chain is different from the reported inhibition effect. The increased proximate particle polarizability caused by increasing particle radius will affect the interaction between the proximate particles and the mains particles, which finally results in an enhancement effect on NFRHT. The dependence of $G$ between two particles on the twisting angle of the proximate ensemble is smooth, and no oscillation exists, which is quite different from the observation in previous work. The proximate MBI will not bring an oscillation of NFRHT when twisting the proximate particle ensemble, which is different from the oscillation phenomenon of NFRHT reported for particle gratings resulting from the intra- and inter-ensemble MBI.

## Acknowledgements

The support of this work by the National Natural Science Foundation of China (No. 51976045, 52206081) and by the 'New Era Longjiang Outstanding Master's and Doctoral Dissertations' project (LJYXL2022-062) are gratefully acknowledged. The support from the China Postdoctoral Science Foundation (2021M700991) is also acknowledged.